\documentclass[draftcls, 12pt,onecolumn,oneside]{IEEEtran}
\usepackage{amsmath}
\usepackage{subfigure}
\usepackage{graphicx}
\usepackage{latexsym}
\usepackage{amssymb,amsbsy,verbatim,array}
\usepackage{epsfig}
\usepackage{cite}
\usepackage{color}
\usepackage{epstopdf}
\usepackage{algorithm}
\usepackage{algorithmic}
\usepackage{psfrag}
\usepackage{wrapfig}

\usepackage{bm}
\usepackage{makecell}
\begin{document}

\title{Deep Reinforcement Learning Based Mobile Edge Computing for Intelligent Internet of Things}
\author{Rui Zhao, Xinjie Wang, Junjuan Xia, and Liseng Fan

\thanks{R. ZHao, J. Xia, and L. Fan are all with the School of Computer Science, Guangzhou University, Guangzhou 510006, China. (email: 2111806073@e.gzhu.edu.cn, xiajunjuan@gzhu.edu.cn, lsfan2019@126.com). }
\thanks{This work was supported in part by the NSFC under Grant 61871139}
\thanks{X. Wang is with Qingdao University of Technology, Qingdao 266520, China. (email: xinjie1023@163.com).}
%\thanks{D. Deng is with the School of Information Engineering, Guangzhou Panyu Polytechnic, Guangzhou, 511483, China
%(e-mail: dengdan@ustc.edu).}
%\thanks{F. Zhu is with Guangdong New Generation Communication and Network Innovative Institute (GDCNi), Guangzhou, China}
%\thanks{The authors declare that there is no conflict of interest regarding the publication of this paper.}
%\thanks{The authors state the data availability in this manuscript through the email to the corresponding author.}
%%\thanks{}
}

%\author{
%\uppercase{Rui Zhao}\authorrefmark{1}, \uppercase{Xinjie Wang}\authorrefmark{2}, \uppercase{Zou Dan}\authorrefmark{3}, \uppercase{Junjuan Xia}\authorrefmark{1}, and \uppercase{Liseng Fan}\authorrefmark{1}}
%\address[1]{School of Computer Science and Cyber Engineering, Guangzhou University, Guangzhou 510006, China}
%\address[2]{Qingdao University of Technology, Qingdao 266520, China}
%\address[3]{School of Information Engineering, East China Jiaotong University, Nanchang, 330013, China}
%\tfootnote{This work was supported in part by the Graduate Innovative Research Grant Program of Guangzhou University under Grant 2019GDJC-M18, and by the Science and Technology Program of Guangzhou under Grant 201807010103.
%}
%\markboth
%{Rui Zhao \headeretal: Deep Reinforcement Learning Based Mobile Edge Computing for Intelligent Internet of Things}
%{Rui Zhao\headeretal: Deep Reinforcement Learning Based Mobile Edge Computing for Intelligent Internet of Things}
%
%\corresp{Corresponding authors:  Junjuan Xia and Liseng Fan (e-mail: xiajunjuan@gzhu.edu.cn, lsfan2019@126.com).}

 \pagestyle{headings}
\maketitle \thispagestyle{empty}

\begin{abstract}
In this paper, we investigate mobile edge computing (MEC) networks for intelligent internet of things (IoT), where multiple users have some computational tasks assisted by multiple computational access points (CAPs). By offloading some tasks to the CAPs, the system performance can be improved through reducing the latency and energy consumption, which are the two important metrics of interest in the MEC networks. We devise the system by proposing the offloading strategy intelligently through the deep reinforcement learning algorithm. In this algorithm, Deep Q-Network is used to automatically learn the  offloading decision in order to optimize the system performance, and a neural network (NN) is trained to predict the offloading action, where the training data is generated from the environmental system. Moreover, we employ the bandwidth allocation in order to optimize the wireless spectrum for the links between the users and CAPs, where several bandwidth allocation schemes are proposed. In further, we use the CAP selection in order to choose one best CAP to assist the computational tasks from the users. Simulation results are finally presented to show the effectiveness of the proposed reinforcement learning offloading strategy. In particular, the system cost of latency and energy consumption can be reduced significantly by the proposed deep reinforcement learning based algorithm.
\end{abstract}

%\begin{IEEEkeywords}
%Deep reinforcement learning, mobile edge computing, intelligent IoT.
%\end{IEEEkeywords}
%
%\titlepgskip=-15pt

\maketitle
\section{Introduction}
In recent years, there has been a great progress in the development and application of wireless communication systems \cite{NanzhaoCommMag2016,NanzhaoTVT2018}, and many new techniques have been proposed to speed up the data rate of wireless communication. Among these techniques, relaying technique is one of the most promising techniques to enhance the communication quality, and it can work in many protocols such as decode-and-forward (DF) and amplify-and-forward (AF). In addition, cognitive technique is also very attractive \cite{ZhaoCJE,ZhaoTVM}, since it can help utilize the spectrum resources very effectively \cite{ZhaoChinacomm,ZhaoTVT,ZhaoTVT2}. Moreover, multiple antenna technique can help enhance the transmission data rate \cite{KOF1,KOF2}, and its newest form of massive  multi-input multi-output (MIMO), which can help improve the transmission data rate by ten or hundred times \cite{ZhongcaijunJSTSP}. %In further, intelligent reflecting surface has attracted much attention from the researches, which depends on the development of communications and material science \cite{Pan1,Pan2}.

With the development and allocation of wireless communication systems, especially about the fifth-generation (5G) networks, there has been a great progress in the development of internet of things (IoT), which can also support the development of smart cities. In IoT systems, the nodes can not only communicate with each other, but also have the ability to store the data and compute. Among the IoT systems, the technique of wireless caching is quite important, since it can help improve the user's experience quality substantially \cite{Gui2020wc,EuraSipXJJ}. Fortunately, the storage cost has been decreasing very rapidly due to the development of storage technique. Besides the wireless caching technique, the technique of mobile edge computing (MEC) plays a very important role in the IoT-based systems \cite{ZichaoTII,GuoMECAccess,ZichaoAccess}, where the nodes can compute the tasks assisted by the near-by nodes instead of remote cloud. In this way, the latency and energy consumption can be reduced substantially.

Driven by the development of big data and deep learning, there has been a trend in the development of intelligent systems, such as the intelligent IoT. In \cite{KeHeTVT,KeheAccess}, the deep convolutional neural networks (CNNs) were incorporated into the conventional detectors such as the maximum likelihood detector (MLD), zero-forcing (ZF), and minimum mean square error (MMSE) detectors, and it can be found that the detection performance can be improved significantly.  In \cite{Gui2020TVT,GuiG2020TVT},  shows a good application of machine learning in flight delay prediction and basic image analysis.  In \cite{Wang2020JESTCS}, deep learning (DL) has been introduced into automatic modulation classification (AMC) due to its outstanding identification performance.  In \cite{Lichao2018}, the Q-learning based intelligent algorithms have been proposed to protect the communication from the smart attacker, which can operate in spoofing, eavesdropping, interfering and silent modes. In \cite{EuraSipChaoLi,Wan11,Wan12,Wan13,ShiweiAccess}, the authors extended to study the intelligent secure algorithms for the wireless communication systems in many application scenarios, such as the non-orthogonal multiple-access (NOMA) systems, imperfect channel estimation and multiple levels of primary users in cognitive networks. In \cite{2016Mastering,Wang2018Prefrontal}, the deep reinforcement learning were incorporated into the strategy game such as  the Weiqi   and it can be found that the improved the winning rate of the machine. In particular, in 2016, ``Alpha Go'' adopted a deep reinforcement learning framework to defeat human go players.

In this paper, we study MEC networks for intelligent IoT, where multiple users have some computational tasks assisted by multiple computational access points (CAPs). By offloading some tasks to the CAPs, the system performance can be improved through reducing the latency and energy consumption, which are the two important metrics of interest in the MEC networks. We devise the system by proposing the offloading strategy intelligently through the deep reinforcement learning algorithm. In this algorithm, deep Q-network is used to automatically learn the  offloading decision in order to optimize the system performance, and a neural network (NN) is trained to predict the offloading action, where the training data is generated from the environmental system. Moreover, we employ the bandwidth allocation in order to optimize the wireless spectrum for the links between the users and CAPs, where several bandwidth allocation schemes are proposed. In further, we use the CAP selection in order to choose one best CAP to assist the computational tasks from the users. Simulation results are finally presented to show the effectiveness of the proposed reinforcement learning offloading strategy. In particular, the system cost of latency and energy consumption can be reduced significantly by the proposed deep reinforcement learning based algorithm.

The organization of this paper is given as follows. After the introduction in this section, we will discuss the system model of MEC networks as well as the linearly weighted cost in Sec. II. Then, we introduce how to intelligently optimize the system performance by using the DQN as well as the bandwidth allocation and CAP selection in Sec. III. Sec. IV will present the simulation results and conclusions are finally made in Sec. V.

\section{system model}

\begin{figure}[tbp]
        \centering
        \includegraphics[width=3.0in]{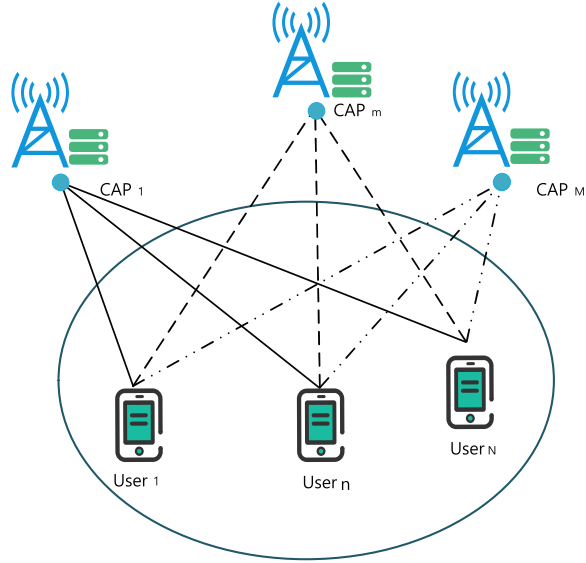}\vspace{-3mm}
        \caption{System model of MEC network with multiple users.}\vspace{-5mm}
        \label{system_model}
        \end{figure}
In the paper, we consider the problem of offloading strategy design for MEC network, in order to determine how many tasks to computed by the CAPs. To further enhance the system performance, the bandwidth allocation is studied to optimize the wireless bandwidth among users and CAPs.  Moreover, we consider the problem of CAP selection to choose one best CAP among multiple ones to assist the computation. {Specifically, the system model is shown in Fig. \ref{system_model}, where we consider a task offloading network with $N$ users $\{u_n|n=1, 2, \dots, N\}$ and $M$ CAP nodes $\{e_m|m=1, 2, \dots, M\}$. All users have only one antenna while the MEC nodes have multiple antennas. For each user $u_n$, we assume that the computational task $\{l_n|n=1,2, \dots, N\}$ can be arbitrarily divided into two parts: one part to be computed at local while the other part to be offloaded to the CAP}. In Fig. \ref{system_model}, the user  $u_n$  firstly selects an optimal MEC node to offload the task $l_n$ through the channel parameters. After that, the user $u_n$ determines an offloading strategy, give by

\begin{align}
	\bm{\alpha}_n = [\alpha_{n,1}, \alpha_{n,2}, \dots, \alpha_{n,m}, \dots, \alpha_{n,M}],
\end{align}

where $m \in \{1, \dots, M\}$ and the corresponding component $\alpha _{n, m} \in [0, 1]$ represents the percentage of the task $l_n$ to be offloaded to the MEC node $e_m$. Since the users only select one MEC node to offload tasks, there is at most one element greater than zero in the offloading strategy vector $\bm{\alpha}_n$. We denote the offloading ratio as
\begin{align}
A_n=\sum_{m=1}^M{\alpha _{n, m}}.
\end{align}
Note that the offloading strategy $\bm{\alpha} _{n}$ includes the following three offloading scenarios:
\begin{enumerate}[1)]
\item $A_n=0$, In this scenario  the task $l_n$ is computed at local.
\item $A_n>0$, In this scenario $A_n$ percents of the task $l_n$ is offloaded to the MEC node $e_m$ while the rest ($1-A_n$)percents of task is computed at local.
\item $A_n = 1$, In this scenari the task $l_n$ is computed at the CAP node  $e_m$.
\end{enumerate}

We assume that the channels follow Rayleigh flat fading. Then, the transmission data rate from the  $u_n$  to MEC node $e_m$ can be obtain from the Shannon theory \cite{Wan7}as
%and the transmission data rate from the user $u_n$ to the MEC node $e_m$ can be obtained by
\begin{align} \label{E1}
    C_{n,m}=W_n \log_2\Big(1 + \frac{P_{tran}^{n}|h_{n,m}|^2}{{\sigma}^2}\Big),
\end{align}
where $W_n$ is bandwidth of the wirless $u_n$-$e_m$ link, $h_{n,m}$ $\mathcal{CN}(0, 2)$  denotes the channel gain of the $u_n$-$e_m$ link,  $P_{tran }^{n}$ represents the transmit power at the user $u_n$, and $\sigma^2$ is the variance of the additive white Gaussian noise (AEGN) at the CAP nodes.

\subsection{Local-Computing Model}
We denote the computing capability (i.e., number of CPU cycles per second) at the user $u_n$ as $f_n$. Then, the local computation time  can be obtain from the  \cite{ZichaoTII} as
 %from \cite{ZichaoTII} which we can obtain the local computation time
\begin{align} \label{E2}
    T_{local}^{n}&=\frac{l_n}{f_{n,m}} (1-{\alpha _{n,m}})\ {\gamma} \ {\omega},
\end{align}
where $\gamma$ is the conversion coefficient from million bits to bits, and $\omega$ is the number of cycles required for the CPU to compute per bit of task. In addition, the local computational energy consumption can be obtain from the  \cite{ZichaoTII} as
\begin{align} \label{E3}
    E_{Local}^{n}&= {T_{local}^{n}} P_{local}^{n},
\end{align}
where $P_{local}^{n}$ is the computational power at the user $u_n$.

\subsection{Computing-Offloading Model}
The transmission time consumed  for the offloading link from $u_n$ to $e_m$ can be described as
\begin{align} \label{E4}
    T_{tran}^{n}&=\frac{l_n}{C_{n,m}} {\alpha _{n,m}}\ {\gamma} \ {\omega}.
\end{align}
Similarly, the transmission energy consumption for the user $u_n$ can be described as
\begin{align} \label{E5}
    E_{tran}^{n}= T_{local}^{T} P_{tran}^{n},
\end{align}
where $P_{tran}^{n}$ denotes the transmit power at the user $u_n$. Since the CAPs generally have steady energies, the computational consumption can be ignored for the CAPs. Morover, the computation time at the CAP node $e^m$ can be denoted by
\begin{align} \label{E6}
    T_{e}^{m}= \frac{l_n}{F_{n,m}} {\alpha _{n,m}}\ {\gamma} \ {\omega},
\end{align}
where $F_{n, m}$ denotes the computational capacity allocated for the user $u_n$ by the MEC node $e_m$. Since the size of the transmitted data returned is small enough, feedback latency and energy consumption can be ignored to simplify the problem. Therefore, we formulate the total system latency as
\begin{align}\label{E7}
    T_{total}   &= \sum_{n=1}^{N}  \left (T_{local}^{n} + T_{tran} ^{n} + T_{e}^{m}     \right).
\end{align}
In addition the total energy consumption is formulated as
\begin{align}\label{E7}
     E_{total} &= \sum_{n=1}^{N}  \left (E_{local}^{n} + E_{tran}^{n}      \right).
\end{align}
Note that minimizing both the total latency and the total energy consumption is a multiple objective optimization problem, which is however very complicated to implement in practice. To simplify the problem, and facilitate theoretical analysis, we consider a linear weighted objective function instead, which is given by
\begin{align}\label{E15}
    \Phi _m  = \lambda T_{total}  +  (1-\lambda) E_{total},
\end{align}
where $\lambda \in{[0, 1]}$ is a weight factor. In short, the optimization problem of minimizing total latency and the total energy  consumption is expressed as

\begin{align}\label{E16}
    \min_{\{ {\alpha _{n,m}},W_n\}} \Phi_m   &\\ \nonumber
    \quad \quad \mathrm{s.t.} \quad
    \text{$C_1$: } & {\alpha _{n,m}} \in [0,1] \\
    \text{$C_2$: } & \sum_{n \in \cal{N}} W_n = W_{total}, \nonumber\\
\end{align}

\section{System Optimization}

The main objective of this paper is to minimize the weighted cost $\Phi_m$. In order to achieve this goal,   we firstly optimize the bandwidth allocation , and then we optimize the offloading strategy based on the allocated bandwidth. Finally, we give the CAP selection strategy based on the results of offloading strategy and bandwidth allocation.

\subsection{Bandwidth Allocation Optimization}
In this part, we employ three bandwidth allocation criteria to assist the users  and meanwhile to reduce the system weighted cost $\Phi _m$.  The simplest bandwidth allocation strategy is the uniform allocation, which has the lowest computational complexity. In this criterion,  $W_n$ is identical for each user $u_n$, given by

\begin{align}
    \label{bandwidth_allocation_1}
    W_n=\frac{W_{total}}{N}.
\end{align}

This bandwidth allocation strategy is independent of specific channel conditions and task length.  In order to incorporate the task length in to the bandwidth allocation,
%match the bandwidth allocation with the tasks of users,
 we present bandwidth allocation criterion \uppercase\expandafter{\romannumeral2} as,
\begin{align}\label{bandwidth_allocation_2}
    W_n = \frac{l_n} {L} W_{total},
\end{align}
which indicates that the allocated bandwidth $W_n$ depends on the task length of users.

Besides these two criteria, we future consider another dynamic bandwidth allocation strategy, which is related to the task offloading. Let i denote the number of iteration in the offloading process, and we use $\alpha _{n,m}^{i}$ to represent the updated  task allocation strategy. Based on $\alpha _{n,m}^{i}$, we obtain a dynamic bandwidth allocation strategy as,
%To future reduce the system total cost $\Phi_m $, we present bandwidth allocation criterion  \uppercase\expandafter{\romannumeral3}. During the task offloading decision process, we will generate a new task allocation strategy $\alpha _{n,m}^{i}$ and the weighted cost $\Phi ^{i}$ at each iteration. Based on this, we consider the dynamic change of bandwidth as well as the allocated bandwidth at each iteration, as follows

\begin{align}\label{bandwidth_allocation_3}
    W_n = \frac {\alpha _{n,m}^{i}} {\sum_{n=1}^{N} \alpha _{n,m}^{i}}  W_{total}
\end{align}
%where $i$ denotes the number of iterations.

\subsection{DQN-Based Resource Allocation}
After allocating bandwidth for a given CAP node, we will continue to optimize the offload strategy to reduce the system cost $\Phi_m $.  Due to the complexity of resource allocation and task scheduling in MEC networks, it is hard to apply the traditional optimization methods solve this problem. Fortunately, the recent reinforcement learning technology has shown good results in solving mobile strategy problems, and  it can be  recognized as an ideal technology to optimize task offloading strategy in MEC networks.
\subsubsection{RL}
\begin{figure}[tbp]
         \centering
         \includegraphics[width=3.0in]{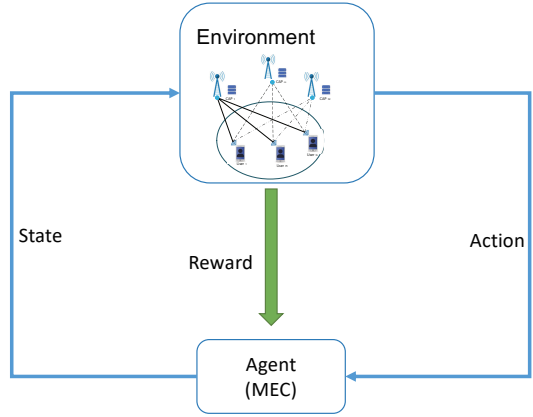}\vspace{-3mm}
         \caption{Reinforcement learning model.}\vspace{-5mm}
         \label{RL_model}
\end{figure}
As shown in Fig. \ref{RL_model}, the framework of reinforcement learning consists of an agent and the corresponding environment which the agent interacting. In this scenario, each MEC node is viewed as an agent and everything except the MEC nodes is regarded as the environment. The agent makes decisions by observing the change of states.
Reinforcement learning is an unsupervised learning technology, for which the agent can find the optimal behavior sequence  via on-line learning. The agent repeatedly interacts with the environment by trial and error. Eventually, the agent modifies its own strategy to adapt to different environments to accomplish tasks.

%We design different reward functions for different decision problems,
In RL, it is of vital importance to design a proper reward function the specific decision problem, with the purpose to reward appropriate behaviors with respect to the current statey.  In this paper, we design the reward function as,
%In the paper,  the reward function we designed mainly gives the difference between the cost$\Phi_{t} $ of the current action and the cost $\Phi_{t-1} $ of the last action.
\begin{align}
    r_=\left\{
           \begin{array}{ll}
              1  , & \hbox{if ( $\Phi_{t} $ - $\Phi_{t-1} $ ) is larger than zero}\\
              -1, & \hbox{other}
           \end{array}
         \right.,
\end{align}
where $\Phi_{t} $ and $\Phi_{t-1} $  denote the total cost at time slot $t$ and $t-1$, respectively.

The task offloading process can be regarded as a Markov decision process(MDP) for during the time domain.  Let $\textbf{S}= \{l_1 \alpha _{1,m}, l_2 \alpha _{2,m},\ldots ,l_N \alpha _{N,m} \}$ be the state space and $\textbf{A}= \{\alpha _{1,m}, \alpha _{2,m},\ldots ,\alpha _{N,m} \}$  be the action space. The set of feasible actions for the state $S_t \in \textbf{S} $ is $A_{S}^{t} $, which is a subset of \textbf{A}.  The transformation from $S_t$ to $S_{t+1} $ with an specific action $A_t$  follows the probability $P(S_{t+1}|S_t , A_t )$.  The action is decided by a policy $\pi$ : $S \rightarrow A$. The policy is obtained by training the agent through reinforcement learning.

\subsubsection{DQN}
\begin{figure}[tbp]
         \centering
         \includegraphics[width=3.0in]{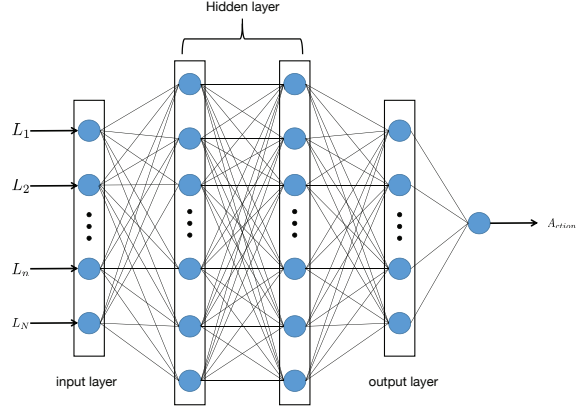}\vspace{-3mm}
         \caption{Structure of the deep neural network.}\vspace{-5mm}
         \label{NN_model}
\end{figure}

The traditional reinforcement methods such as Q-learning and Sarsa learning have a common feature, which is to use tables to store  state value functions. However, for the general task offloading problem, the value function cannot be saved in the form of table due to huge state dimensions. Therefore, we choose DQN to solve the task offloading problem. Compared with Q-learning, DQN uses a deep neural network(DNN) with parameters $ \omega $ as a value function approximator to solve the task offloading problem. As shown in Fig. \ref{NN_model}, we take state $s$ as the input of the DNN.  The DNN consists of an input layer $H^i$, the $k$ shared hidden layers $\{H_1,H_2,..., H_k\}$, and an input layer $H_{j}^{i}$. In order to get the next action, we set a greedy selection strategy on the output of the DNN, with probability $\epsilon$ to select a random action $a$, which can be described as follows,

\begin{align}
    A=\left\{
           \begin{array}{ll}
              random , & \hbox{when the probability is $\epsilon$ }\\
              \mathop{\arg\min}_{a}(Q(s_t,a;\omega)) , & \hbox{when the probability is $1 - \epsilon$}
           \end{array}
         \right .
\end{align}

There is a certain correlation between the interaction sequence and the state action in RL. If we train the neural network directly, the effect of the model will not be good as expected. To solve this problem, we adopted the experience replay structure proposed by DeepMind team in NeurIP in 2013.

The replay buffer includes two parts: the collecting samples and the sampling samples. The collected samples are stored in replay buffer according to the time sequence. If the replay buffer is full of samples, them the new samples will overwrite the oldest sample in time sequence. Generally speaking, a batch of samples will be randomly sampled from the cache evenly for learning, and the training effect will be more stable. At the same time, a sample will be trained many times to improve the sample utilization rate.

Note that the traditional reinforcement learning updates the status value based on the return value of the current time and the next estimated value. But the instability of the data causes the neural network training results to fluctuate at each iteration. These fluctuations will be reflected at the next iteration, and hence the training  result is difficult to be stable. In order to deal with the deviation in temporal difference and reduce the impact of correlation, we need to decouple the two parts as much as possible. Hence, we introduce the target network. Firstly, the two models use the same parameters before the training. Secondly, during the training process, behavior network is responsible for interacting with the environment and getting interaction samples. Then, in the learning process, the target value is calculated by target network, and the target value is obtained by comparing the estimated values of the target network and then the behavior network, and the behavior network is updated. Finally, when the iterations reach a certain number, the parameters of the behavior network are synchronized to the target network,  and the next stage of learning can be carried out. Similar to the supervised learning, we define the loss function of the DQN as the variance between the target value $Q_{targe} $ and the predicted value $Q(s,a:\omega)$ weights $\omega$ to minimize the loss,

\begin{align}
    Loss_{\omega}&= ((r -\gamma {\arg\min}_{a}(Q(s',a';\omega '))) - Q(s,a:\omega))^2
\end{align}

Therefore, the target value is fixed in a certain period of time through the target network and the target network reduces the volatility of the model eventually.

The process of DQN algorithm is described in algorithm 1. The main steps are as follows.
\begin{enumerate}[]
\item  Using a neural network with a parameter of $\omega$ as the approximator of $Q $ value.
\item  Defining a loss function using the mean square error of the $Q$ value.
\item  Calculating the gradient of loss function for the parameter $\omega$.
\item Using Stochastic Gradient Descent(SGD) to optimize the parameters.
\end{enumerate}

\begin{algorithm}[tbp]
  \caption{DQN-Based MEC network Resource Allocation Optimization}
  \label{DQN_algor}
{\textbf{Input:}} {user task $l$,radio bandwidth resource $W_{total}$, computing capability  of users and CAP nodes}\\
{\textbf{Output:}} {task offloading result a}\\
{Initialize replay memory D to capacity N}\\
{Initialize action-value function Q with weight $\omega$}\\
{Initialize tager action-value function $\widehat{Q} $ with weights $\omega '=\omega$}
{initialize states $s_1$}
  \begin{algorithmic}
  \FOR{t = 1,T}
  \STATE {with probability $\omega$ select a random action $a_t$}
  \STATE {$
    a_t=\left\{
           \begin{array}{ll}
              random , & \hbox{ $\epsilon$ }\\
              \mathop{\arg\min}_{a}(Q(s_t,a;\omega)) , & \hbox{$1 - \epsilon$}
           \end{array}
         \right.$
}
    \STATE Perform $a_t$ in the environment, observe the reward r and the next state $s_{t+1}$
    \STATE  Store transition ($s_t$,$a_t$,$r_t$,$s_{t+1}$) in D
    \STATE Sample random minibatch of  transitions  ($s_i$,$a_i$,$r_i$,$s_{i+1}$) from D
    \STATE $y_i = r_i + {\arg\min}_{a}(\widehat{Q} (s_t',a';\omega ')$
    \STATE Performing gradient descent
    \STATE Interval C step to update $\widehat{Q}$ = Q

  \ENDFOR

\end{algorithmic}
\end{algorithm}

\subsection{CAP Selection}
After optimization of bandwidth allocation and offloading, in order to further reduce the system cost $ \Phi _m $ we performed a CAP selection operation. In this work, we propose a method for choosing the best CAP to help users perform calculations.

For this MEC network, the user firstly sends some pilot signals, from which the MEC estimates the associated channel parameters.  Then, the channel which has the smallest gain is measured among the N channels,
\begin{align} \label{RANASH}
  \theta_m = \min_{m \in [1,M]} \{\vert h_{1,m}\vert^2,  \vert h_{2,m}\vert^2,...,\vert h_{N,m}\vert^2\}.
\end{align}

Each CAP is associated with of $\theta_m $. Fron the set $\{\theta_m| 1 \leq m \leq M   \}$, we select one best CAP which has the largest $\theta_m $ among M ones,
%Then, the channel which has the largest parameter is selected from the selected $M$ channels $ \theta_m $, and the corresponding MEC is optimal with respect to the selected channel, the best MEC node is given by,

\begin{align} \label{eq:relaysectionCriterionII}
  e^* = \arg  \mathop {\max }\limits \theta_m .
\end{align}

\section{Simulation Results}
In this part, we use several bandwidth allocation strategies and CAP selection methods to evaluate the proposed optimization algorithm. All channels in the network experience Rayleigh flat fading. If not specified, the transmit and computing powers at the users are set to 2W and 3W, respectively. The deep neural network has two hidden layers. The CPU of each CAP has the same computing power with computational capacity of $6.3\times 10^8$ cycle per second (cyc/sec) . Moreover, the  six users have different computational capacities, which are  set to $1.4\times10^8$ cyc/sec,  $0.21\times10^8$ cyc/sec,  $0.95\times10^8$ cyc/sec,  $0.13\times10^8$ cyc/sec,$0.53\times10^8$ cyc/sec and $0.52\times10^{8}$ cyc/sec, respectively. The task sizes of the six users are set to 5.3Mb,3.5Mb, 4.6Mb,3.0Mb, and 4.2Mb, respectively. In further, the total bandwidth of the wireless links is set to 10MHz, so that $B_{total}=10$MHz.

\begin{figure}[tbp]
    \centering
    \includegraphics[width=4.0in]{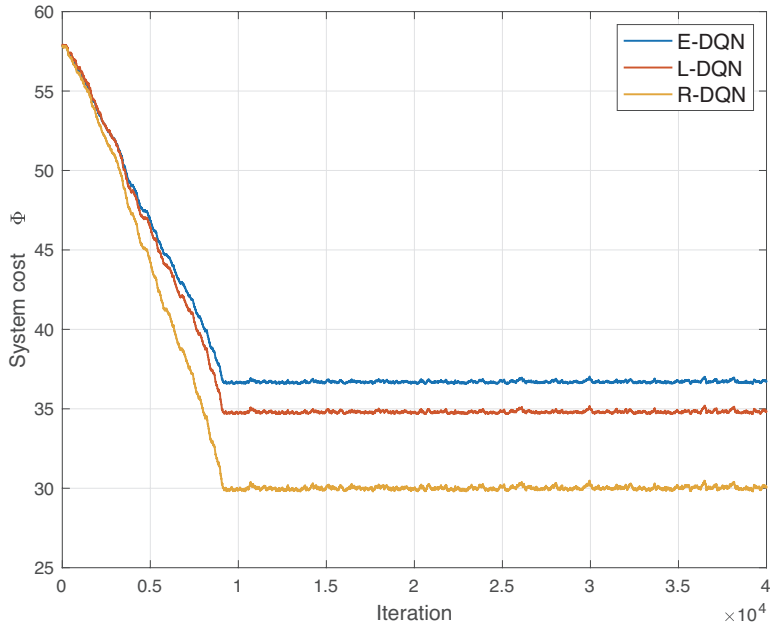}
    \caption{Convergence of the DQN algorithm versus iteration.}
    \label{DQN1}
\end{figure}

Fig. \ref{DQN1} shows the convergence of the proposed DQN algorithm, where M=2, N=5, and $B_{total}$=10MHz.  For the convinced of notation, we used ``R-DQN'',  ``L-DQN'',  and  ``E-DQN''  to represent the DQN with the bandwidth allocation based on the offloading ratio in the iteration, bandwidth allocation by the sub-task length and equal bandwidth allocation, respectively. From (\ref{DQN1}), we can see that the DQN with several bandwidth allocation schemes converge swiftly, and after 8000 iterations, system can achieve stable performance. Moreover, the performance of L-DQN is better than that of E-DQN, as L-DQN in corporates the length of sub-tasks into the bandwidth allocation process. In further, we can find that R-DQN outperforms E-DQN and L-DQN, indicating that the offloading ration in the iterative process can help allocation the wireless bandwidth very effectively.
\begin{figure}[tbp]
         \centering
         \includegraphics[width=4.0in]{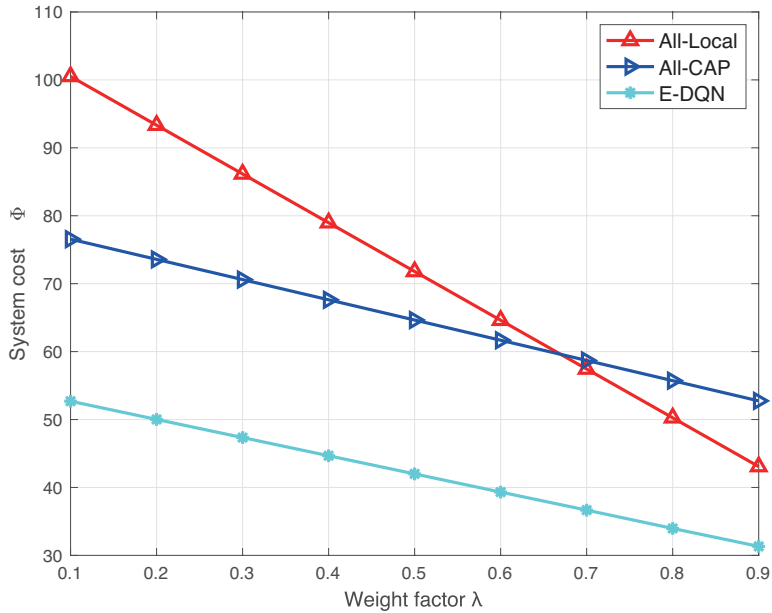}
         \caption{Comparison of the three  offloading strategies   versus
the weight factor $\lambda$.}
         \label{lambda_cost_tl}
\end{figure}

In Fig. \ref{lambda_cost_tl}, we show the relationship between the system cost  $\Phi$ and several offloading strategies and the weight factors $\lambda$, where M=2 ,N=5, and  $\lambda$ varies from 0.1 to 0.9. In the Fig. \ref{lambda_cost_tl} we use 'All-Local' and 'All-CAP' to represent that the tasks are computed locally and by the CAPs, respectively. The equal bandwidth allocation scheme is adopted in Fig. \ref{lambda_cost_tl}. From this figure, we can find that the proposed E-DQN outperform the 'All-Local' and 'All-CAP' for various values of  $\lambda$, indicating that the proposed scheme can efficiently utilize the computational resources among the users and CAPs. Moreover, the cost of 'All-CAP' is smaller then that of 'All-Local' when $\lambda$ is small, as using the CAPs to compute the tasks can help reduce the energy consumption. On the contrary, when $\lambda$   is large, 'All-CAP' becomes course them 'All-Local', simple the transmission latency becomes the bottle neck of the system cost.

\begin{figure}[tbp]
         \centering
         \includegraphics[width=4in]{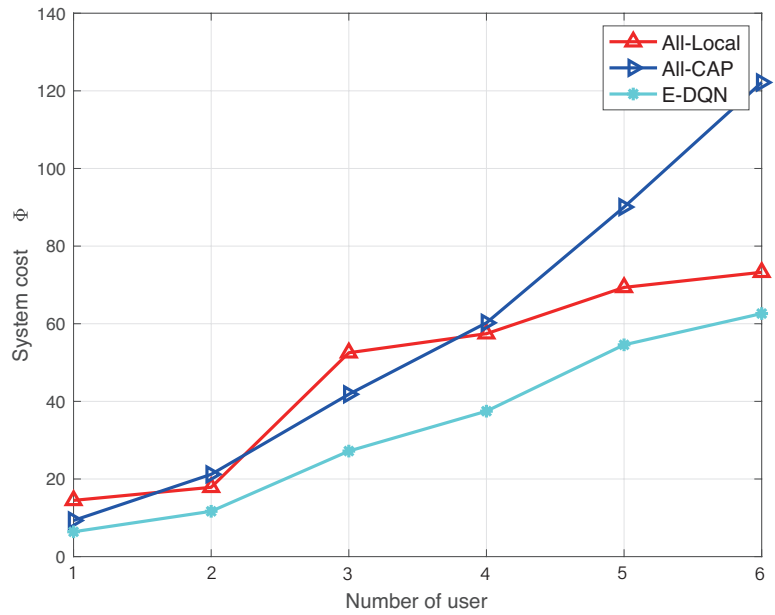}
         \caption{ Comparison of the three  offloading strategies   versus
the number of users.}
         \label{Number_user}
\end{figure}
Fig.  \ref{Number_user} shows the impact of number users on the system cost with several offloading strategies, where M=2 and N varies from 1 to 6. For performance comparison, we present the cost of the proposed 'E-DQN', 'All-Local' and 'All-CAP' in this figure. From Fig.  \ref{Number_user}, we can find that the system costs increases with a larger value of N, as more users give more burden of the computational tasks on the system. Moreover, for various values of N, the proposed 'E-DQN' outperforms the 'All-Local' and 'All-CAP', which further validates the effectiveness of the proposed scheme in scheduling the computational resources in the system.

%Fig.  \ref{Number_user} show the impact of changes in the number of users on system weights under three different offloading strategies. From these figure,  We can see that with the increase in the number of users, the weight of the system is also increasing. Among them, E-DQN is still the best among the three strategies, which verifies the effectiveness of the DQN algorithm.

\begin{figure}[tbp]
         \centering
         \includegraphics[width=4in]{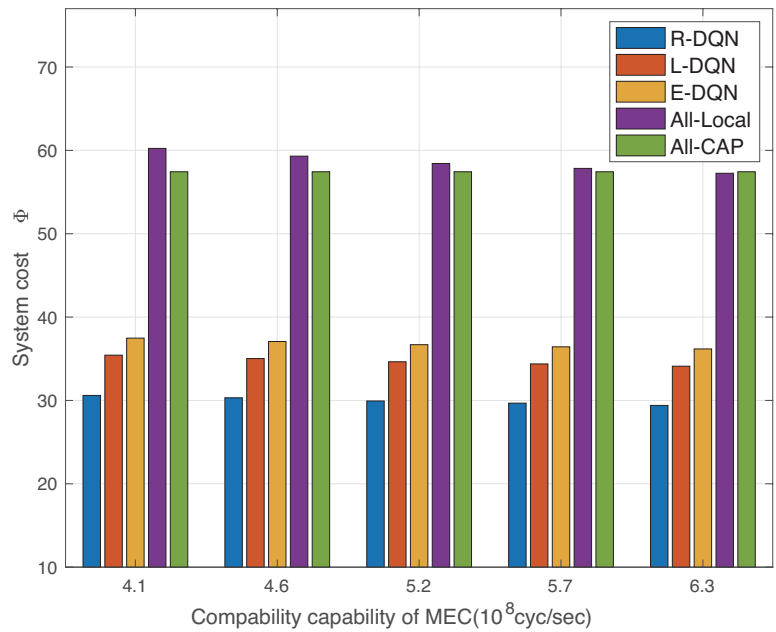}
         \caption{The computing capability of MEC }
         \label{different_computing}
\end{figure}

Then, we investigate the impact of the computational capability of the CAP on the cost. From Fig. \ref{different_computing}, we can observe that the maximum and minimum CPU  frequency of the set to CAP are $6.3\times10^8$ cyc/sed and $4.1\times10^8$ cyc/se, respectively.
The performance of 'All-Local' computing and 'All-CAP' computing are still the worst. We can observe that when E-DQN, E-DQN or E-DQN algorithm is adopted, the average energy consumption of the system decreases when the computing capability of the servers increases, because these three schemes can upload more tasks to the CAPs. The results indicate  that offloading tasks to the CAP server can be completed faster, thereby reducing the cost. We can also observe that user's offloading rates increase as the computational  power of the edge server increases. This indicates that when the user finds that the computing resources of the CAP are sufficient, the user is more willing to upload more tasks to the CAP.

\begin{figure}[tbp]
         \centering
         \includegraphics[width=4in]{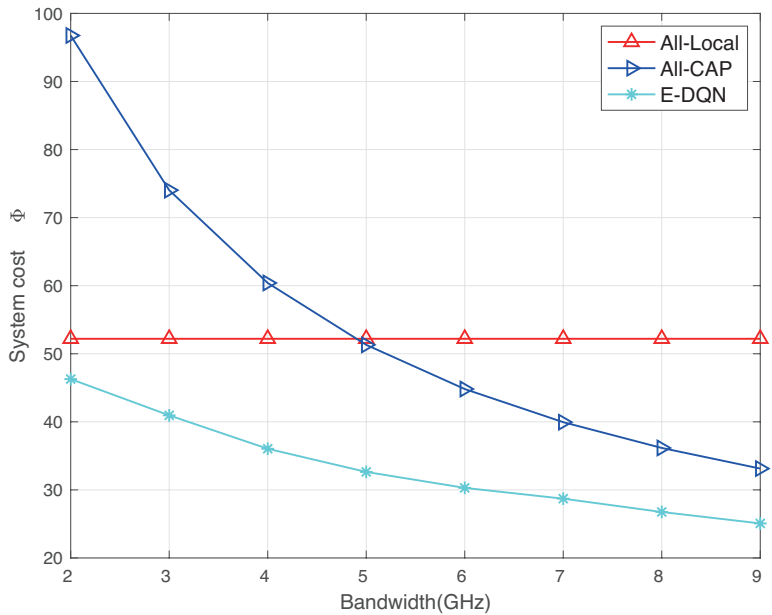}
         \caption{Comparison of the three  offloading strategies   versus
the bandwidth $W_{total}$}
         \label{different_bandwidth}
\end{figure}

 In Fig. \ref{different_bandwidth}, we compare the cost performances of the several offloading strategy versus the wireless bandwidth, where M=2, N=5, and the bandwidth $B_{total}$ varies from 2GHz to 9GHz. We can find from Fig. \ref{different_bandwidth} that for various values of $B_{total}$, the proposed E-DQN scheme outperforms  the 'All-Local' and 'All-CAP', which  further validates the effectiveness of the proposed offloading strategy. Moreover, unlike the 'All-local', the proposed E-DQN and 'All-CAP' have better performances when the value of $B_{total}$ becomes larger. This is because greater bandwidth can help reduce transmission delays and transmission energy consumption.

\begin{figure}[tbp]
         \centering
         \includegraphics[width=4in]{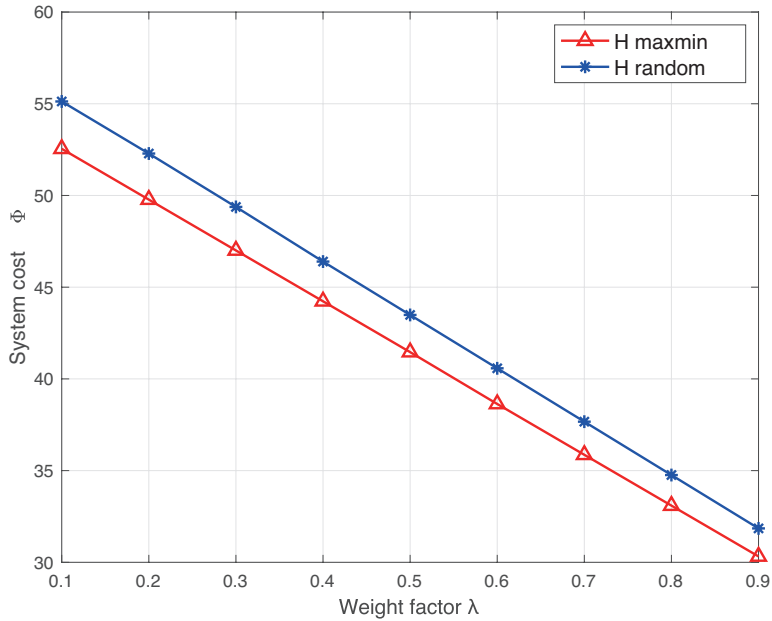}
         \caption{Comparison of the CAP selection method with E-DQN versus the weight factor  $\lambda$.}
         \label{CAP_selection}
\end{figure}
In Fig.  \ref{CAP_selection}, we show the effect of CAP selection on the system cost performance versus the weight factor $\lambda$, where M=2, N=5, and $B_{total}=10$MHz. For comparison, we plot the result of random CAP selection in Fig.  \ref{CAP_selection} as a benchmark. As observed from Fig.  \ref{CAP_selection}, we can find that for various of   $\lambda$, the CAP selection scheme outperforms the random selection scheme, since the former can exploit the wireless links for improving the transmission latency and energy consumption. The validity of the proposed research in this work is further verified.

\section{Conclusions}

This paper studied MEC networks for intelligent IoT, where multiple users have some computational tasks assisted by multiple CAPs. We devised the system by proposing the offloading strategy intelligently through the deep reinforcement learning algorithm. In this algorithm, Deep Q-Network was used to automatically learn the  offloading decision in order to optimize the system performance, and a neural network (NN) was trained to predict the offloading action, where the training data was generated from the environmental system.

 Moreover, we employed the bandwidth allocation in order to optimize the wireless spectrum for the links between the users and CAPs, where several bandwidth allocation schemes were proposed. In further, we used the CAP selection in order to choose one best CAP to assist the computational tasks from the users. Simulation results were finally presented to show the effectiveness of the proposed reinforcement learning offloading strategy. In particular, the system cost of latency and energy consumption could be reduced significantly by the proposed deep reinforcement learning based algorithm.

%In further works, we will apply the considered MEC networks into the application of IoT based systems such as the works in \cite{HuijunRenewable2020,HuijunRenewable2019,HuijunSustainable}. Moreover, we will consider to use some other intelligent algorithms \cite{Ding1,Ding2} to the considered system, in order to further enhance the system performance by reducing the latency and energy consumption.

\bibliographystyle{IEEEtran}
\bibliography{IEEEabrv,CRN}

\end{document}